\documentclass[pra,twocolumn,aps,amssymb]{revtex4}
\usepackage{amssymb}

\usepackage{amssymb}

\usepackage{graphicx}
\usepackage{amsmath}
\usepackage{times}

\begin{document}

\title{Effects of finite temperature on the Mott insulator state}

\author{Guido Pupillo$^{1,2,3}$, Carl J. Williams$^{1}$, and Nikolay V. Prokof'ev$^{4,5}$}
\affiliation{$^1$National Institute of Standards and Technology,
Gaithersburg, MD 20899
\\$^2$Department of Physics, University of Maryland, College Park, MD 20742
\\$^3$ Institute for Quantum Optics and Quantum Information,
6020 Innsbruck, Austria
\\$^4$Department of Physics, University of Massachusetts,
Amherst, MA 01003
\\$^5$Russian Research Center ``Kurchatov Institute'', 123182, Moscow}

\begin{abstract}
We investigate the effects of finite temperature on ultracold Bose atoms
confined in an optical lattice plus a parabolic potential
in the Mott insulator state. In particular, we analyze the
temperature dependence of the density distribution of atomic pairs in the lattice,
by means of exact Monte-Carlo simulations.
We introduce a simple model that quantitatively accounts for the
computed pair density distributions at low enough temperatures.
We suggest that the temperature dependence of the atomic pair statistics
may be used to estimate the system's temperature at energies of the
order of the atoms' interaction energy.
\end{abstract}
\maketitle

\section{Introduction}
Great progress has been achieved in the coherent control and
manipulation of ultracold atoms and molecules. In a series of recent
experiments, several groups were able to demonstrate the loading of
an atomic Bose-Einstein condensate into the lowest vibrational level
of single wells of an optical lattice
\cite{Greiner,Peil2003a,Soferle,Inguscio,Fertig}. In a remarkable
experiment, M. Greiner {\it et al.}, \cite{Greiner}, demonstrated
the capability of inducing a reversible quantum phase transition
between a superfluid and an insulator state for bosonic atoms by
varying the intensity of the trapping laser beams and therefore the
depth of single potential wells. In a series of recent experiments,
the same group demonstrated the capability of creating an array of
quasi one dimensional tubes and induce an effective one dimensional
superfluid/insulator transition, with on-site peak density equal to
one \cite{Paredes,Gerbier,Foelling}. Several proposals suggest this
system has potentially fundamental applications in the field of
quantum computation, as the realized array of atoms in the
insulating regime can be used to create an ideal register for qubits
\cite{Jaksch99,Zoller2003,GAG,BHdynamics,Viverit2004}.

A key observable in experiments with ultracold atoms is the
interference pattern observed after a certain time of flight
following the release of atoms from all trapping potentials. The
interference pattern is proportional to the distribution of atomic
momenta \cite{Prokofev,Roberts,Guido,Rigol}, and provides direct
information on the amount of phase coherence present in the system.
Information extracted from the interference pattern is routinely
used to measure the system temperature in the superfluid regime, but
cannot be easily used in the strongly interacting regime, where both
finite temperature and interactions compete to fill up the
single-particle Brillouin zone. In fact, in time-of-flight images
the measurement precision is on the order of the level spacing to
the second lattice band, while the relevant energy scales in the
strongly interacting regime are the onsite interaction energy and
the hopping amplitude. The latter are typically at least one order
of magnitude smaller than the lattice level spacing \cite{Jaksch}.
Thus, apart of a scheme based on fermionization of the many-body
wavefunction \cite{Paredes}, to date there are no available schemes
for the estimation of temperature at energies of the order of the
interaction energy in the strongly correlated regime.

Recently, time-of-flight images have provided evidence for the
mixing of particle-hole excitations in the ground state in the Mott
regime Ref.~\cite{Gerbier,Foelling}. This mixing has important
consequences for the use of the Mott state as a quantum register. In
fact, the presence of empty and doubly occupied sites introduces
unwanted errors in the register initialization, which have to be
corrected for at the cost of overheads in computational resources
and gate times. A characterization of the finite-temperature
population of particle-hole states in realistic experimental
conditions is thus desirable.

In this work we analyze the effects of finite temperature on the
Mott insulator state in the presence of an external parabolic
potential. We focus on the effects of temperature on the density
distribution of doubly occupied sites, showing that at finite
temperature the latter has a characteristic site-dependent Gaussian
profile due to the presence of the external potential. We derive the
functional form of this temperature dependence by introducing an
effective single-particle model for the many-body spectrum at
energies of the order of the interaction energy, and we compare our
model's predictions for the atomic pair density distribution to
exact quantum Monte-Carlo results for realistic experimental
parameters of the trapping potentials, finding good agreement. We
suggest that the temperature dependence of the density distribution
of doubly occupied sites may be used to estimate the system's
temperature at energies of the order of the interaction energy.

The presentation of the results is organized as follows. In
Sec.~\ref{MottState} we review the physics of the Mott insulator
state and the low-energy spectrum of the Bose-Hubbard Hamiltonian
both in homogeneous and inhomogeneous lattices. In
Sec.~\ref{Effective} we introduce an effective model for the
many-body spectrum at energies of the order of the interaction
energy, and provide an analytical expression for the temperature
dependence of the density distribution of atomic pairs at low enough
temperatures. In Sec.~\ref{Simulations} we validate our model by
comparing the analytical results to exact quantum Monte-Carlo
simulations. In Sec.~\ref{Finite-T} we address the stability of the
Mott insulator state against finite temperature population of
particle-hole states. In Sec.~\ref{Estimation} we discuss the
possibility of estimating the system's temperature by collecting
statistics of the presence of atomic pairs in the lattice. Finally,
the conclusions are presented in Sec.~\ref{Conclusions}.

\section{The Mott insulator state}\label{MottState}
The Bose-Hubbard Hamiltonian describes the system's dynamics
when the lattice is loaded such that only the lowest vibrational
level of each lattice site is occupied \cite{Jaksch}
\begin{equation}
H_{BH}=\sum_{j} \left[ \epsilon(j)n_j-J(a_j^{\dagger}a_{j+1}
+a_{j+1}^{\dagger}a_{j})+\frac{U}{2}n_j(n_j-1) \right].
\end{equation}
Here $a_j$ is the bosonic annihilation operator of a particle at
site $j$ and $n_j=a_j^{\dagger}a_{j}$. The site-dependent
$\epsilon(j)$ models an external magnetic quadratic potential, while
$J$ and $U$ are the tunneling and on-site interaction energies
respectively. For deep lattices with potential energy $V(x)=V
cos^2(kx)$, the tunneling energy is approximated by
$J=4/(\sqrt{\pi})E_{R}(V/E_{R})^{3/4}e^{-2\sqrt{V/E_{R}}}$, where
$E_R=(\hbar k)^2/2m$ is the recoil energy, $k$ the light wave
vector, and $m$ the atomic mass. The on-site interaction energy is
due to ground state collisions between atoms each in the motional
state $\phi(\mathbf{x})$ and is given by
$U=\frac{4 \pi a_s\hbar^2}{m}\int d \mathbf{x}|\phi(\mathbf{x})|^4$,
with $a_s$ the s-wave scattering length and $\phi(\mathbf{x})$ a Wannier state.
$J$ and $U$ are assumed to be site independent \cite{Jaksch}.

For a homogeneous one dimensional lattice ($\epsilon(j)=0$, $
\forall j$) of $M$ sites with periodic boundary conditions at zero
temperature $T$, an insulating state, the {\it Mott state}, occurs
only if the number of particles $N$ equals $M$ and the interaction
energy dominates over the tunneling energy. If $\sqrt{N} J/U \ll 1$
the ground state has approximately energy $E_g=-4NJ^2/U$ and is
given by
\begin{equation}
|\Psi_g\rangle=\alpha(|T\rangle+2\sqrt{N}J/U|S\rangle),
\end{equation}
where the normalization constant is $\alpha=(1+4N(J/U)^2)^{-1/2}$.
The Fock state $|T\rangle=\Pi_{j=1}^N a^{\dagger}_j|0\rangle$ has
one particle per site and zero energy, and the symmetrized state
$|S\rangle =1/\sqrt{2 N}\sum_{j=1}^N(a_j^{\dagger}a_{j+1}
+a_{j+1}^{\dagger}a_{j})|T\rangle$ has energy $U$. Indeed,
in the homogeneous system all Fock
states with an empty site and an atomic pair in another site are
degenerate with energy $U$. The latter is roughly the lowest
excitation energy for the many-body system in the homogenous
commensurately filled case \cite{Fisher}.

The probability of creating exactly $N=M$ particles in an experiment
is negligible. An insulator state can be recovered if an external
quadratic potential is present. Then, $\epsilon(j)=\Omega j^2$ where
$\Omega=m(\pi/k)^2\omega_T^2/2$ is a characteristic energy scale for
a trap of frequency $\omega_T$. At $T=0$, a Mott state with average
occupation one in the center of the trap can be realized for
$\sqrt{N} J/U \ll 1$ if
\begin{equation}
N<M,  \;\;\;\;\;\;  U>\epsilon((N-1)/2),  \;\;\;\;\;\;   \Omega
N>J.\label{Req}
\end{equation}
The last two inequalities insure that multiple particle occupancy
and tunneling of holes from the borders to the center of the lattice
are suppressed, respectively \cite{BHdynamics}. The first inequality
in Eq.~\ref{Req} simply states the number of wells be larger than
the number of atoms. As a consequence, there are $M!/((M-N)!N!)$
Fock states with maximal on-site occupation one, at variance with
the single state of the case $N=M$. In the presence of the trap
these states are not degenerate, spanning an energy range which may
be larger than $U$. We redefine as
$|T\rangle=\Pi_{j=-(N-1)/2}^{(N-1)/2} a^{\dagger}_j|0\rangle$ the
lowest energy state and set its energy to zero
\cite{DefinitionLattice}. The state $|T\rangle$ is the ground state
of the system to zeroth order in $J$, while the true ground state
has coherences due to tunneling of a hole in either one of sites
$\pm (N-1)/2$ and mixing of Fock states $|S_{n,m} \rangle=
a^{\dagger}_n a_m |T\rangle/\sqrt{2}$ with an atomic pair and a hole
at sites $n$ and $m$ respectively, with $n,m<|(N-1)/2|$, in analogy
to the homogeneous case. States $|S_{n,m} \rangle$ are not
degenerate, due to the trap presence. States $|S_{n,m} \rangle$ with
$m = n \pm 1$ dominate the mixing into the ground state, and for
$\Omega \ll U$ the amplitude of mixing approaches the homogeneous
system's one $ \sqrt{2} \alpha J/U$.

    In general, the $|T\rangle$-state is coupled {\it via}
the $BH$-Hamiltonian to $|S_{n,m} \rangle$-states with a matrix
element of order $(J/U)^{|n-m|}$, which is negligible for $|n-m|$
large enough in the insulating regime. On the other hand, a
temperature of the order of the energy difference between $|T
\rangle$ and $|S_{n,m} \rangle$-states allows for population of the
latter. In particular, the lowest energy Fock state
$|S_{n,m}\rangle$ has two particles in the central site of the
lattice and a hole at site $|(N-1)/2|$. Its energy is
$\tilde{\Delta}=U-\epsilon((N-1)/2)$. This state is an eigenstate of the
$BH$-Hamiltonian for $J=0$ only. For any finite $J$, eigenstates are
a superposition of many different Fock states. This superposition
lowers the energy $\tilde{\Delta}$ of an amount $W$ which is of the 
order of a few $J$. As a consequence, the energy $\Delta$ of the lowest-energy 
eigenstate with two atoms in a site is reduced to 
approximately $\Delta=\tilde{\Delta}-W$.
 In the following we
introduce a simple model for the spectrum of the many-body system at
energies of order $\Delta$, which we find to be the relevant energy
scale for the double occupation of a site in the lattice \cite{GAG}.

\section{Effective single-particle spectrum}\label{Effective}
Neglecting for a moment mixing of $|S_{n,m}\rangle$-states in low
lying modes, we expect that the density distribution of atomic pairs
as a function of the position in the lattice at finite temperature
$T$ has a a Gaussian profile for the following argument. Given a
lattice with $M$ wells, $N=M+1$ particles and $U \gg J$, the ground
state has roughly one particle per site and an extra particle at the
center of the trap, forming an atomic pair. For weak enough
quadratic traps, $\Omega \ll J$, tunneling of the extra particle is
not suppressed over a certain number of lattice sites at the center
of the trap where the external potential is essentially flat,
meaning that the atomic pair acts approximately as a conventional
harmonic oscillator with {\it effective} mass $m_*=1/(4 J)$ and
trapping frequency $\omega_*= 2 \sqrt{2 \Omega J}$. Here the lattice
constant $a=\pi / k$  and $\hbar = h/2 \pi$ have been set to unity,
where $h$ is the Planck constant. The tight binding spectrum is
therefore approximated by an harmonic oscillator spectrum whose
level spacing is $\omega_*$ with an error of order $O(\Omega/J)$. If
$T > \omega_*/k_B$, it can be shown that the quantum density
distribution $P(j)$ for the atomic pair becomes equivalent to the
classical distribution which is proportional to
\begin{equation}
P(j)\propto e^{- \beta m_* \omega_*^2 j^2 /2} = e^{- \beta \Omega
j^2} \label{Prob1}
\end{equation}
where $\beta = 1/(k_B T)$, and $k_B$ is Boltzmann constant.
The distribution $P(j)$ as a
function of lattice site $j$, Eq.~\ref{Prob1}, is therefore a
Gaussian whose width is $x_0 =\sqrt{1/(2\beta\Omega)}$.\\

We postulate that for realistic situations where $M>N$ harmonic
oscillator states actually approximate the incomplete basis set
defined on the central $N$ sites of the lattice in the energy range
$\Delta \lesssim E \ll 3 U - 2\epsilon((N-1)/2)$, where $3 U -
2\epsilon((N-1)/2)$ is a characteristic energy for population of
states with three particles in one of the central $N-2$ sites. Apart
of a normalization factor, we therefore expect that the density
distribution as a function of the position $j$ in the lattice has
the same Gaussian profile as in Eq.~\ref{Prob1} with an exponential
suppression due to the energy shift $\Delta$
\begin{equation}
P(j) \propto e^{- \beta \Delta} e^{- \beta \Omega j^2}. \label{Prob}
\end{equation}
The width $x_0$ of the density distribution can be therefore
directly related to the temperature of the sample, for $k_B
T\lesssim \Delta$. In fact, while for a temperature of order
$\Delta$ states are populated with more than one pair in the central
sites, the various pairs approximately behave like non-overlapping
harmonic oscillators, meaning that the probability of finding two
extra particles in the same site is strongly suppressed. Therefore,
the density distribution at a single site should still approximately
maintain a Gaussian profile of width $x_0$. Moreover, the height of
the Gaussian peak is an indication of $\Delta$, and therefore of the
number of particles $N$.  
For vanishing values of $J$, $\Delta$ is approximately $\tilde{\Delta}$,
and the peak's height is easily computed. For $J>0$, the shift $W=\Delta-\tilde{\Delta}$
is of the order of half of the pair's band-width.
Because the latter is $8J$, $W$ is approximately $4 J$.
Because the peak's height also depends on the normalization, 
we expect Eq.~\ref{Prob} to well approximate the
value of the density peak for $k_B T \ll \Delta$,
where the normalization is almost one.\\

The modeling of the excited states of the many-body system
with an effective harmonic oscillator spectrum for atomic pairs
is one of the central results of this paper. In the following we verify the accuracy of
this picture by comparing the model's predictions for the
atomic pair density distribution to exact Monte-Carlo results.
\begin{figure}
\begin{center}
\leavevmode {\includegraphics[width=3. in]{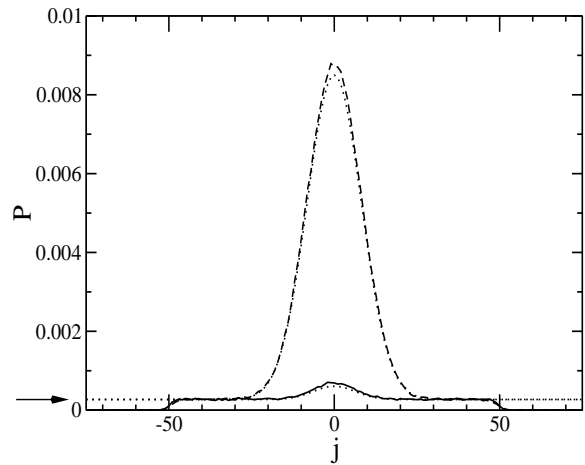}}
\end{center}
\caption{Density distribution of atomic pairs $P$ as a function of
the lattice index $j$. Continuous and dashed lines are numerical
results for $T=3 J/k_B$ and $T= 5 J/k_B$ respectively. Dotted lines
are analytical curves for the same temperatures. The arrow indicates
the zero-temperature mixing of $|S_{n, n\pm 1}\rangle$-states. Here
$N=101$, $U/J=120$, $\Omega/J=0.0374226$, and
$\Delta/J=24$.}\label{Fig1}
\end{figure}

\begin{figure}
\begin{center}
\leavevmode {\includegraphics[width=3. in]{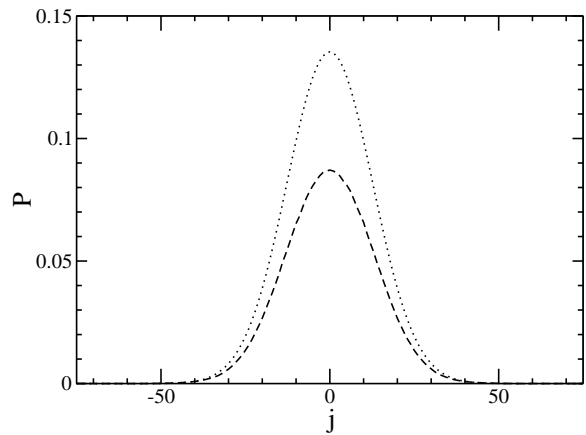}}
\end{center}
\caption{Density distribution of atomic pairs $P$ as a function of
lattice index $j$, for $T=12 J/k_B$. Dashed and dotted lines are
numerical and analytical results respectively. Here $N=101$,
$U/J=120$, $\Omega/J=0.0374226$, and $\Delta/J=24$.\label{Fig2}}
\end{figure}

\section{Numerical simulations}\label{Simulations}
We perform numerical simulations of the density distribution of
atomic pairs in the lattice as a function of temperature for some
realistic experimental parameters. We employ a quantum Monte-Carlo
code based on the continuous-time Worm algorithm \cite{Proko}. A
sample of $N=101$ atoms of $^{87}Rb$ is trapped in a lattice with
wavelength $\lambda=785$ nm and parallel and transverse confinements
$V_{\parallel}=15 E_R$ and $V_{\perp}=40.5 E_R$ respectively. Using
$a_s=5.6$ nm, the interaction energy in frequency units is $U/h =
3.340$ kHz and $U/J=120$. The external magnetic trapping frequency
is $\omega_T\approx 2 \pi \times 40$ Hz and therefore the number of
atoms $N$ fixes $\epsilon{(N-1)/2}\approx 0.8U$, and thus $\Delta \approx 24
J$. 
We have chosen these parameters because they are experimentally
feasible and satisfy Eq.~\ref{Req}, so that at zero temperature a
Mott state is formed with one particle occupying each one of the
central $N$ sites of the lattice. In particular, zero temperature
mixing of basis states with an empty site in one of the central $N$
sites of the lattice is suppressed, the largest amplitude of mixing
corresponding to the outer most trapped particle tunneling to the
adjacent empty site. Calculations have been performed for $M \gg N$,
so that atoms never reach the border of the lattice. We note that
$N=101$ compares well with the typical number of atoms present in
the one-dimensional tubes of
Refs.~\cite{Peil2003a,Soferle,Inguscio,Fertig,Paredes,Gerbier,Foelling}.
For example, $N$ is equal to 80 in the central tube of
Ref.~\cite{Fertig}, which is the largest number of atoms for the
entire array of tubes, while $N$ is as small as 20 in the central
tube of Ref.~\cite{Paredes} .\\

\begin{figure}
\begin{center}
\leavevmode {\includegraphics[width=3. in]{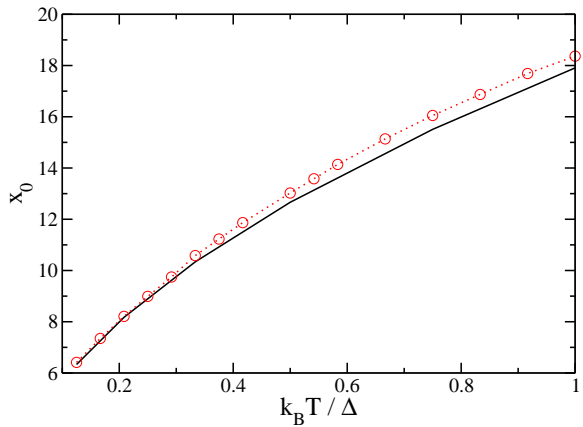}}
\end{center}
\caption{(Color online) Width $x_0$ of the density distribution of
atomic pairs as a function of $k_B T/\Delta$. The width $x_0$ is in
units of the lattice constant $a$. Circles(red) are numerical
values, while the continuous line is the analytic curve $(k_B T/ 2
\Omega )^{1/2}$. Here $N=101$, $U/J=120$, $\Omega/J=0.0374226$, and
$\Delta/J=24$.}\label{Fig3}
\end{figure}
In Fig.~\ref{Fig1} the atomic pair density distribution $P$ is
plotted as a function of the site index $j$ for $T k_B /J=3$ and
$5$. Continuous and dashed lines are numerical results, while dotted
lines are analytical curves. For $T k_B /J=3$, lower curves, the
numerical solution shows essentially a flat density distribution
throughout the central $N$ sites, with a shallow gaussian peak at
the center. The flat distribution corresponds to the zero
temperature residual mixing of $|S_{n,n\pm 1}\rangle$-states into
the ground state, characteristics of the Mott state. The results of
Ref.~\cite{Gerbier} are consistent with the observation of this
mixing. As $\Omega \ll U$, corrections to the density distribution
due to the external trapping potential are not distinguishable on
the scale of the graph, and the on-site density matches the
homogeneous system's value $2 \times 2(J/U)^2 \approx 0.000278$. The
factor of two in front of last expression is due to the fact that
the extra atom can tunnel from the left or right. The latter
constant has been added to the analytic curves, and is indicated by
an arrow in Fig.~\ref{Fig1}. Finite atomic pair density due to
zero-temperature mixing of $|S_{n, n\pm 1}\rangle$-states is a
direct signature of the creation of a Mott state. For sufficiently
low temperatures, selective probe of the density distribution in
lattice sites $j$ with $j\gg x_0$ can give direct {\it in situ}
evidence of the formation of the Mott plateau in the center of the
trap, while the shallow peak around $j=0$ measures finite
temperature population of $|S_{n,m}\rangle$-states.
\begin{figure}

\begin{center}

\leavevmode {\includegraphics[width=3.5 in]{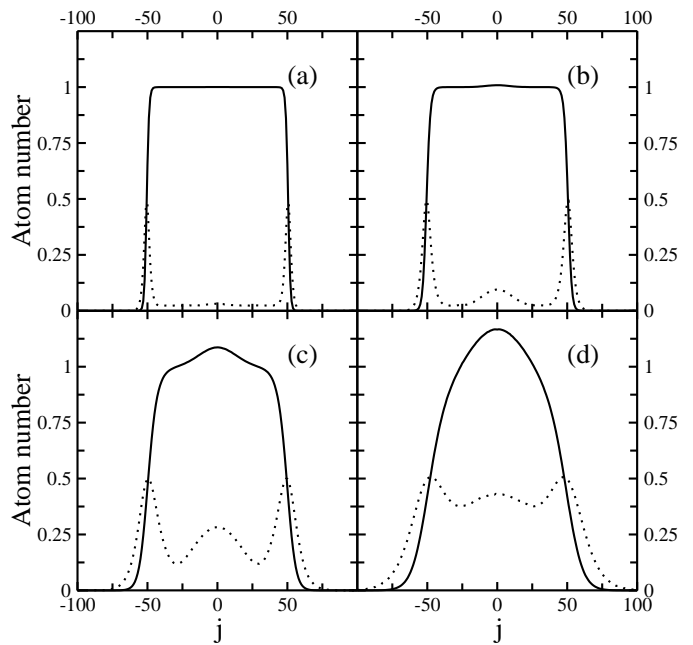}}
\end{center}

\caption{Local densities and fluctuations as a function of the site
index $j$ for $k_B T / J=3(a),5(b),12(c),$ and $24(d)$. Here,
$\Delta/J=24$. Continuous and dotted lines are the numerical
densities and fluctuations, respectively.}\label{Fig4}
\end{figure}

 For $T k_B /J=5$
the Gaussian peak is more evident on the scale of the graph. We
observe that numerical and analytical curves nearly overlap. In
particular, the widths of the Gaussians perfectly match, while the
height of the peak is slightly underestimated by the analytic curve.
This is due to the fact that $\Delta$ is only
approximately $24 J$ for the chosen parameters, and the difference
is amplified by the exponential function.
A larger disagreement in the peak height is observable for $k_B
T/J=12$ in Fig.~\ref{Fig2}, where the analytic curve largely
overestimates the exact result. This is expected, as for higher
temperatures the normalization constant differs more and more from
one, therefore suppressing the peak height in the exact solution.

In Fig.~\ref{Fig3} the width $x_0$ is plotted against the
temperature, up to $T=\Delta/k_B$. Circles are numerical results,
while the continuous line is the analytic solution $\sqrt{k_B T/(2
\Omega)}$. The plot shows an almost perfect agreement between
numerical and analytical results for $k_B T \lesssim \Delta/4$.
Deviation for higher temperatures is due to population of states
with a large number of atomic pairs. For $k_B T=\Delta$ the
deviation is still less than a lattice constant. We consider the
good agreement between the analytic model and the numerical results
over the entire range $k_B T\leq \Delta$ as
a strong indication of the validity of the harmonic oscillator model.\\

\section{Finite-T depletion of the Mott state}\label{Finite-T}
A large population of states with two particles in a site will
ultimately totally deplete the Mott state, thus making any
application of the Mott insulator state to quantum computation
schemes impossible. In the remainder of this section we investigate
the role of the energy scale $\Delta$ in the thermal depletion of
the Mott insulator state, by numerically analyzing the atomic
density profile and onsite number fluctuations.

In Fig.~\ref{Fig4} the on-site atomic density $\langle n_j \rangle$
(continuous lines) and the number fluctuations $\langle \Delta n_j
\rangle=\sqrt{\langle n_j^2 \rangle - \langle n_j \rangle^2}$
(dotted lines) are plotted as a function of the site index $j$, for
$k_B T / J=3(a),5(b),12(c),$ and $24(d)$. For $k_B T / J=3$ the
density profile is flat, which corresponds to an extended Mott state
around $j=0$. The number fluctuations are also flat throughout the
lattice, except for two large peaks around $\pm N/2$ and a small
bump at the trap center. The flat regions correspond to the
zero-temperature mixing of particle-hole states
\cite{Guido,Ana2005}, which is approximately given by $2 \sqrt{2}
J/U$, while the large peaks are due to the tunneling of particles to
unoccupied sites, and are equal to $\Omega N$ \cite{Guido,Ana2005}.
The small bump at the trap center corresponds to the
finite-temperature population of high-energy states with two
particles in a site, analogous to the discussion of Fig.~\ref{Fig1}.
Panels $(b)$ and $(c)$ show that for higher temperatures atoms tend
to accumulate at the trap center, therefore decreasing the extension
of the region of constant density. For $k_B T= 24 J = \Delta$
the Mott state has completely disappeared.

In Ref.~\cite{DeMarco} the effects of finite temperature on the
stability of the Mott insulator state have been studied by means of
classical Monte-Carlo simulations, finding that well defined Mott
insulating plateaux are present for temperatures at most of the
order of $U/(10 k_B)$. The discussion above shows that, for any
fixed $U$ and $\Omega$, this is true only in the limit of a small
number of particles $N$. In fact, we showed that the relevant energy
scale for population of doubly occupied site is
$\Delta=U-\Omega((N-1)/2)^2$, which is $N$-dependent. In particular,
for a system with $N=110$ instead of $N=101$, we have $\Delta\approx
9 J$. In this case, a temperature of the order of $U/(10 k_B)=12
J/k_B$ would be larger than $\Delta$ and would therefore completely
deplete the Mott insulator state.

\section{Temperature estimation}\label{Estimation}

As mentioned above, it is notoriously difficult to measure
temperatures of the order of the interaction energy in the strongly
interacting limit.

\begin{figure}[b]
\begin{center}

\leavevmode {\includegraphics[width=2.7 in]{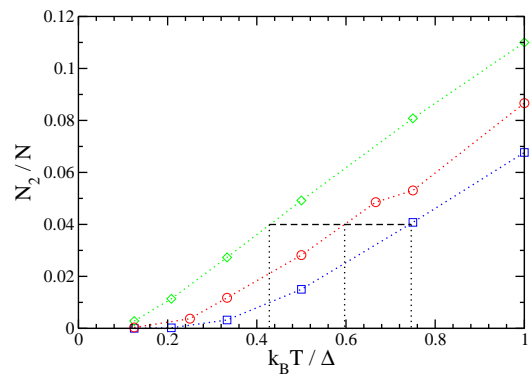}}
\end{center}

\caption{(Color online) Number of atomic pairs $N_2$ as a function
of $k_B T/\Delta$. Here, $N_2$ is in units of $N=101$. The
squares(blue), circles(red), and diamonds(green) correspond to
systems with $N_{tot}=95,101$, and 107 atoms, respectively. For all
curves, $\Delta/J=24$. The dashed line is the error bar for the
temperature estimation, given a measured number of pairs equal to 5
($N_2/N\approx 0.04$). The dotted lines are a guide to the
eye.}\label{Fig5}
\end{figure}

On the other hand, we have shown above that the atomic pair density
distribution has a clear temperature dependence. Here we suggest
that in principle this temperature dependence may be used to
estimate the system's temperature at energies of the order of
$\Delta$, by devising a probe that is sensitive to the presence of
multiply occupied sites in the lattice. For this purpose, two
different kinds of experiments may be considered. The first
possibility is to design an experiment which is solely sensitive to
the total number of pairs in the lattice. The latter can be
calculated numerically as a function of temperature by means of
quantum Monte-Carlo simulations, and the results of the simulations
can then be compared to the experimental data to extract a
temperature estimation.
An example of these calculations for the trapping parameters given
above and $N=N_{tot}=101$ is given in Fig.~\ref{Fig5}, where the
total number of pairs $N_2$ is plotted as a function of the
temperature. In the plot, the circles (red) are the quantum
Monte-Carlo results for $N_{tot}=101$, while the squares (blue)
 and the diamonds (green) correspond to a total number of particles
$N_{tot}=95,$ and $107$, respectively. Each curve is normalized to
$N=101$. The curves for $N_{tot}=95,107$ are shown in order to take
into account a plausible experimental uncertainty on the total
number of atoms $N_{tot}$. In fact, it is experimentally challenging
to control the total number of particles $N_{tot}$ to better than 5
percent, and the uncertainty on the total number of particles is
likely to dominate the experimental errors \cite{PaulLett}. Then,
assuming unit efficiency of atomic pair detection, for each measured
$N_2$ the temperature is easily estimated. For example, for the
system considered above, where $N_{tot}=101$, if 5 pairs
($N_2/N\approx 0.04$) are measured experimentally, a simple analysis
of Fig.~\ref{Fig5} yields a most-probable temperature of about 14.4
$J/k_B$ (0.6 $\Delta/k_B$). Thus, by taking into account the
uncertainty on the total number of atoms (black-dashed line in the
plot), the final estimate for the temperature is $T=(14.4\pm 4.0)
J/k_B$.

We note that the discussion above refers to the determination of
temperature in a single experiment. If one wants to determine
average system's temperatures, the measurement can be repeated many
times, and the error bars are obtained by assuming a shot-to-shot
Poissonian fluctuation of $N_{tot}$, in analogy to the discussion
above. However, in the remainder of this section we discuss an
experiment that allows for the estimation of average system's
temperatures and it is largely independent of the fluctuations of
$N_{tot}$ (although it is admittedly difficult to implement).

In principle various schemes may be used to experimentally detect
the presence of pairs in the lattice. Here, we propose to collect
statistics of atomic pairs through resonant photo-association of two
atoms in a single well to a molecular excited state, followed by
ionization of the formed molecules and detection of the emitted
ions. In fact the latter can be performed with almost unit
efficiency \cite{PaulLett}. For this experiment, the molecular
excited state should be chosen to be far from atomic dissociation,
meaning that the atomic probability of spontaneous emission is not
enhanced by the photo-associative laser.
\\

The second kind of experiment that may be envisioned is one where
the probe has a high spatial resolution, so that it is possible to
determine the position of the atomic pair in the lattice. In this
case, information on average system's temperatures may be extracted
by measuring the width $x_0$, which has been shown above to depend
only on the temperature and the known trap geometry. The advantage
of this experiment is that $x_0$ is insensitive to $N_{tot}$. Thus,
calibration of average system's temperatures may be performed by
accumulating statistics of pair detection on successive experiments
with the same trapping potentials, without worrying about possible
fluctuations of $N_{tot}$ from one experiment to the next.

Although a careful discussion of the sources of uncertainty on the
measurements is out of the scope of the present analysis, we expect
that a significant uncertainty on the temperature measurement is
likely to come from the finite spatial resolution of the
probe-field. In fact, for typical experimental setups it is not
possible to focus a photo-associative laser onto a single site. The
intensity of the photo-associative laser has a Gaussian profile
whose width $\sigma_L$ is typically of the order of the light's
wavelength or larger. This means that the laser intensity may not be
negligible on a few lattice sites, depending on the ratio between
the wavelength of the photo-associative laser and the lattice
constant $a$. A convolution between the pair density-distribution
Gaussian and the laser-intensity Gaussian provides an estimate of
the experimentally measured width $x_0'$, which is thus given by
$x_0'=\sqrt{x_0^2+\sigma^2_L}$. We note that the difference between
$x_0$ and $x_0'$ decreases with increasing temperature. For example,
for a reasonable value $\sigma_L=3a$, $x_0'-x_0$ is approximately
$0.6$ $a$ for $k_B T=0.2 \Delta$, while it drops to $x_0'-x_0\approx
0.3a$ for $k_B T=0.6 \Delta$.\\

Finally, we note that it is straightforward to generalize the
calculations above for the array of one-dimensional tubes of
Refs.~\cite{Peil2003a,Soferle,Inguscio,Fertig,Paredes,Gerbier,Foelling}.
Assuming an initial Thomas-Fermi distribution of the atoms in the
three dimensional system, the calculations above may be repeated for
each tube separately and the results combined (see also
Ref.~\cite{Paredes}).

\section{Conclusions}\label{Conclusions}
In summary, we have studied the effects of temperature on the Mott
insulator state, with an emphasis on the finite temperature
population of states with two atoms in a site. We introduced a model
for the energy spectrum of the trapped many-body system in the
strongly correlated regime, which allowed us to derive the
functional form of the dependence of the density of atomic pairs
upon temperature. We discussed the depletion of the Mott state due
to the population of states with multiple on-site occupancy.
Finally, we suggested that the temperature dependence of the density
distribution of doubly occupied sites may be used to estimate the
system's temperature at energies of the order of the interaction
energy.

{\it Note added:} After this work was completed, the results of an
experiment analogous to one of those proposed here have been
presented in Ref.~\cite{Esslinger1}, where the temperature of a
fermionic Mott state has been estimated.

\section{Acknowledgements}
The authors thank Gavin K. Brennen, Andrea Simoni, Boris V.
Svistunov, Eite Tiesinga, Paul Lett, and Trey Porto for useful
discussions. This research was supported in part by ARDA/NSA.


\begin{thebibliography}{99}
\bibitem{DefinitionLattice} For convenience we choose $N$ odd and
fix the site index $j=0$ at the minimum of the magnetic trap.
\end{thebibliography}
\end{document}